\documentclass[12pt]{article}
\usepackage{epsfig}
\usepackage{array}
\usepackage{pennames}
\usepackage{rotating}

\newcommand\pubnumber{OREXP-01-01}
\newcommand\pubdate{\today}
\newcommand\hepnumber{hep-ph/0105076}

\def\csumb{Department of Physics\\
University of Oregon, Eugene, OR 97403 USA}
\def\support{\footnote{Work supported by the
US Department of Energy grant DE-FG03-96ER40969.}} 

\def\Title#1{\begin{center} {\Large\bf #1 } \end{center}}
\def\Author#1{\begin{center}{ \sc #1} \end{center}}
\def\Address#1{\begin{center}{ \it #1} \end{center}}

\newcommand\pubblock{\rightline{\begin{tabular}{l} \pubnumber\\
         \pubdate\\ \hepnumber \end{tabular}}}
\newenvironment{Abstract}{\begin{quotation}  }{\end{quotation}}
\newenvironment{Presented}{\begin{quotation} \begin{center} 
             Presented at the\end{center}
      \begin{center}\begin{large}}{\end{large}\end{center} \end{quotation}}
\def\Acknowledgments{\bigskip  \bigskip \begin{center}
          \large\bf Acknowledgments\end{center}}

\makeatletter
\def\section{\@startsection{section}{0}{\z@}{5.5ex plus .5ex minus
 1.5ex}{2.3ex plus .2ex}{\large\bf}}
\def\subsection{\@startsection{subsection}{1}{\z@}{3.5ex plus .5ex minus
 1.5ex}{1.3ex plus .2ex}{\normalsize\bf}}
\def\subsubsection{\@startsection{subsubsection}{2}{\z@}{-3.5ex plus
-1ex minus  -.2ex}{2.3ex plus .2ex}{\normalsize\sl}}

\renewcommand{\@makecaption}[2]{%
   \vskip 10pt
   \setbox\@tempboxa\hbox{\small #1: #2}
   \ifdim \wd\@tempboxa >\hsize     
       \small #1: #2\par          
     \else                        
       \hbox to\hsize{\hfil\box\@tempboxa\hfil}
   \fi}

 \def\citenum#1{{\def\@cite##1##2{##1}\cite{#1}}}
 
\newcount\@tempcntc
\def\@citex[#1]#2{\if@filesw\immediate\write\@auxout{\string\citation{#2}}\fi
  \@tempcnta\z@\@tempcntb\m@ne\def\@citea{}\@cite{\@for\@citeb:=#2\do
    {\@ifundefined
       {b@\@citeb}{\@citeo\@tempcntb\m@ne\@citea\def\@citea{,}{\bf ?}\@warning
       {Citation `\@citeb' on page \thepage \space undefined}}%
    {\setbox\z@\hbox{\global\@tempcntc0\csname b@\@citeb\endcsname\relax}%
     \ifnum\@tempcntc=\z@ \@citeo\@tempcntb\m@ne
       \@citea\def\@citea{,}\hbox{\csname b@\@citeb\endcsname}%
     \else
      \advance\@tempcntb\@ne
      \ifnum\@tempcntb=\@tempcntc
      \else\advance\@tempcntb\m@ne\@citeo
      \@tempcnta\@tempcntc\@tempcntb\@tempcntc\fi\fi}}\@citeo}{#1}}
\def\@citeo{\ifnum\@tempcnta>\@tempcntb\else\@citea\def\@citea{,}%
  \ifnum\@tempcnta=\@tempcntb\the\@tempcnta\else
  {\advance\@tempcnta\@ne\ifnum\@tempcnta=\@tempcntb \else\def\@citea{--}\fi
    \advance\@tempcnta\m@ne\the\@tempcnta\@citea\the\@tempcntb}\fi\fi}
\makeatother

\newcommand{\inmath}[1] {\ifmmode#1\else$#1$\fi}
\newcommand{\definmath}[2] {\def#1{\ifmmode#2\else$#2$\fi}}
\newcommand{\gVf}  {g_{\mathrm{Vf}}}
\newcommand{\gAf}  {g_{\mathrm{Af}}}
\newcommand{\gVl}  {g_{\mathrm{V}\ell}}
\newcommand{\gAl}  {g_{\mathrm{A}\ell}}

\newcommand{\alphas} {\alpha_{\mathrm{s}}}

\newcommand{\GF}  {G_{\mathrm{F}}}
\newcommand{\PZ}   {\mbox{$\mathrm{Z}$}}  
\definmath{\PWpm} {\mathrm{W}^{\pm}}      
\definmath{\Plp} {\ell^{+}}        
\definmath{\Plm} {\ell^{-}}        
\definmath{\Plpm}   {\ell^{\pm}}         
\definmath{\Pgtp} {\tau^{+}}        
\definmath{\Pgtm} {\tau^{-}}        
\definmath{\Pgtpm}   {\tau^{\pm}}         
\definmath{\Pgn}  {\nu}          
\definmath{\Pagn} {\overline{\nu}}     
\definmath{\Pf}      {\mathrm{f}}
\definmath{\Paf}  {\overline{\mathrm{f}}}
\definmath{\Pq}      {\mathrm{q}}
\definmath{\Paq}  {\overline{\mathrm{q}}}
\definmath{\Pu}      {\mathrm{u}}
\definmath{\Pau}  {\overline{\mathrm{u}}}
\definmath{\Pd}      {\mathrm{d}}
\definmath{\Pad}  {\overline{\mathrm{d}}}
\definmath{\Ps}      {\mathrm{s}}
\definmath{\Pas}  {\overline{\mathrm{s}}}
\definmath{\Pc}      {\mathrm{c}}
\definmath{\Pac}  {\overline{\mathrm{c}}}
\definmath{\Pb}      {\mathrm{b}}
\definmath{\Pab}  {\overline{\mathrm{b}}}
\definmath{\Pt}      {\mathrm{t}}
\definmath{\Pat}  {\overline{\mathrm{t}}}
\definmath{\Pap}  {\overline{\mathrm{p}}}
\definmath{\Pan}  {\overline{\mathrm{n}}}
\definmath{\PaD}  {\overline{\mathrm{D}}}
\definmath{\PaDz} {\overline{\mathrm{D}}^{0}}
\definmath{\PaB}  {\overline{\mathrm{B}}}
\definmath{\PaBz} {\overline{\mathrm{B}}^{0}}
\definmath{\PsDpm}   {\mathrm{D}^{\pm}_{\mathrm{s}}}  
\definmath{\PcgLpm}  {\Lambda^{\pm}_{\mathrm{c}}}  
\definmath{\PD} {\mathrm{D}}     
\definmath{\PDst} {\mathrm{D}^{*}}     
\definmath{\PgLz} {\Lambda^{0}}        

\newcommand{\massof}[1] {m_{\smash{#1}\mathstrut}}

\newcommand{\mHiggs} {\massof{\mathrm{Higgs}}}
\newcommand{\mh}     {\massof{\mathrm{h}}}
\newcommand{\mt}   {\massof{\mathrm{t}}}

\newcommand{\mPZ} {\massof{\mathrm{Z}}}

\newcommand{\GZ}     {\Gamma_{\mathrm{Z}}}
\newcommand{\sighadpole}   {\sigma_{\mathrm{had}}^{\mathrm{pole}}}

\newcommand{\Rl}  {R_{\ell}}
\newcommand{\Rb}  {\mbox{$R_{\rm b}$}}
\newcommand{\Rc}  {R_{\rm c}}
\newcommand{\AFBpole}   {A_{\mathrm{FB}}^0}

\newcommand{\AFBb}   {A_{\mathrm{FB}}^{\mathrm{0,b}}}
\newcommand{\AFBc}   {A_{\mathrm{FB}}^{\mathrm{0,c}}}
\newcommand{\AFBe}   {A_{\mathrm{FB}}^{\mathrm{0,e}}}

\newcommand{\epem}   {\Pep\Pem}

\newcommand{\ff}     {\Pf\Paf}

\newcommand{\qqbar}  {\Pq\Paq}

\newcommand{\ccbar}  {\Pc\Pac}
\newcommand{\bbbar}  {\Pb\Pab}

\newcommand{\Gammaof}[1]   {\Gamma_{\!\smash{#1}\mathstrut}}

\newcommand{\Ghad}      {\Gammaof{\mathrm{had}}}
\newcommand{\roots} {\sqrt{s}}

\newcommand{\pz}  {p_{z}}

\definmath{\GeV}  {\mathrm{GeV}}
\definmath{\GeVc} {\mathrm{GeV}\!/c}
\definmath{\GeVcc}   {\mathrm{GeV}\!/c^2}
\definmath{\MeV}  {\mathrm{MeV}}
\definmath{\MeVc} {\mathrm{MeV}\!/c}
\definmath{\MeVcc}   {\mathrm{MeV}\!/c^2}
\definmath{\MVm}  {\mathrm{MV}\!/\mathrm{m}}
\definmath{\keV}  {\mathrm{keV}}
\definmath{\keVcm}   {\mathrm{keV}\!/\mathrm{cm}}
\definmath{\kV}      {\mathrm{kV}}
\definmath{\km}      {\mathrm{km}}
\definmath{\meter}   {\mathrm{m}}
\definmath{\cm}      {\mathrm{cm}}
\definmath{\mm}      {\mathrm{mm}}
\definmath{\micron}  {\mu\mathrm{m}}
\definmath{\nm}      {\mathrm{nm}}
\definmath{\kg}      {\mathrm{kg}}
\definmath{\gram} {\mathrm{g}}
\definmath{\second}  {\mathrm{s}}
\definmath{\microsec}   {\mu\mathrm{s}}
\definmath{\degree}  {^\circ}
\definmath{\degC} {^\circ\mathrm{C}}
\definmath{\ohm}  {\Omega}
\definmath{\Mohm} {\mathrm{M}\Omega}
\definmath{\rad}  {\mathrm{rad}}
\definmath{\mrad} {\mathrm{mrad}}
\definmath{\nb}      {\mathrm{nb}}
\definmath{\pb}      {\mathrm{pb}}
\newcommand{\eqref}[1]  {(\ref{#1})}
\newcolumntype{L} {>{$}l<{$}}
\newcolumntype{C} {>{$}c<{$}}
\newcolumntype{R} {>{$}r<{$}}

\newcommand {\thw}        {\theta_{\mathrm{W}}}
\newcommand {\thweff}     {\theta_{\mathrm{W}}^{eff}}

\newcommand {\theffl}     {\theta_{\mathrm{eff}}^{\mathrm{lept}}}
\newcommand {\swsq}       {\sin^2\!\thw}
\newcommand {\swsqweff}   {\sin^2\!\thweff}

\newcommand {\swsqeffl}   {\sin^2\!\theffl}

\begin{document}
\begin{titlepage}
\pubblock

\vfill
\def\thefootnote{\fnsymbol{footnote}}
\Title{ Electroweak Measurements on the $\PZ$~Resonance }
\vfill
\Author{David Strom \support}
\Address{\csumb}
\vfill
\begin{Abstract}
Almost all precision electroweak measurements from
the $\PZ$ resonance made at the LEP storage ring by
the ALEPH, DELPHI, L3 and OPAL experiments and those
made using a polarized electron beam at the SLC by the
SLD experiment are now final and have been published.
Changes in the measurements since the last RADCOR
meeting are discussed.  The internal consistency of
the measurements is considered.  The impact of remaining
theoretical uncertainties in the QCD sector are 
examined as well as the impact of experimental
and theoretical uncertainties
on the value of $\alpha_{\mathrm{QED}}(m_{\mathrm{Z}})$.

\end{Abstract}
\vfill
\begin{Presented}
5th International Symposium on Radiative Corrections \\ 
(RADCOR--2000) \\[4pt]
Carmel CA, USA, 11--15 September, 2000
\end{Presented}
\vfill
\end{titlepage}
\def\thefootnote{\arabic{footnote}}
\setcounter{footnote}{0}

\section{Introduction}
Between 1989 and 1995 the LEP collaborations collected
more than 200 pb$^{-1}$ per experiment,
which resulted in
more than 17 million recorded $\PZ$ decays.  
The LEP data included almost 50 pb$^{-1}$ of off-peak
data, which is needed to determine the mass
and width of the $\PZ$. 
The SLD at the SLC recorded collisions with a polarized
electron beam and an unpolarized positron beam
from 1992 until 1998, 
achieving
electron polarizations as high 80\% 
and a total data sample of more
than 500,000 recorded 
$\PZ$ decays.  
These large data samples provide the
basis for well-known tests of the Standard Model.

Almost all of the precision measurements made at
the $\PZ$ resonance have now been published and
the LEP and SLD electroweak working groups have
almost completed a comprehensive review of these measurements
which will shortly appear in {\em Physics Reports}\cite{bib:phr}.
Furthermore, several very complete reviews of the 
theory necessary for the interpretation of these
measurements already exist, see, for example, References 
\cite{bib:montagna,bib:precision,bib:ewbook}.
Rather than attempting to summarize this entire work 
in a few pages, I will briefly review
changes in theory and
measurement since the last RADCOR meeting and then
consider three areas where some controversy
exists:  the determination of strong coupling
constant, $\alphas$, the possible discrepancy 
between the Standard Model and measurements of 
forward-backward asymmetries in 
$\bbbar$ final states, and finally the impact 
on Higgs mass limits of
the contribution
of  hadronic vacuum polarization to the running value
of the electromagnetic coupling constant at the $\PZ$ resonance, 
$\alpha(\mPZ^2)$.

\def\tableline{\noalign{
\hrule height.7pt depth0pt\vskip3pt}}

\begin{table}[t!]

\def\pz{\phantom{0}}
\def\pzz{\phantom{00}}
\newcommand{\mcc}[1]{\multicolumn{1}{c|}{#1}}
\caption[]{\label{tab:all}
  The summary of measurements included in the combined analysis of
  Standard Model parameters used by the LEP and SLD electroweak
  groups reproduced from References~\cite{bib:lepew2000} and
  \cite{bib:lepew1998}.
  The electroweak measurements from $\Pp\bar{\Pp}$
  colliders, $\nu$N scattering, and 
  LEP2 $m_{\PW}$ are described elsewhere in the RADCOR
  2000 proceedings~\cite{bib:tevatron,bib:lep2}.
}

{\small
\begin{center}
\renewcommand{\arraystretch}{1.10}
\begin{tabular}{|l||r|r|r|r|}
\hline
 & \mcc{2000 Result}  &\mcc{1998 Result} & \mcc{Standard} & \mcc{2000} \\
 & \mcc{ }            &\mcc{}            & \mcc{Model fit}&  \mcc{Pull} \\
\hline
\hline
&&&& \\[-3mm]
 $\Delta\alpha^{(5)}_{\mathrm{had}}(m_{\PZ}^2)$
                & $0.02804 \pm 0.00065$ 
                & $0.02804 \pm 0.00065$  &0.02804& $ 0.0$ \\
&&&& \\[-3mm]
\hline
 \underline{LEP}     &&&& \\
 
$m_{\PZ}$ [\GeV{}] & $91.1875\pm0.0021\pz$
                & $91.1867\pm0.0021\pz$
&91.1874$\pz$ & $ 0.0$ \\
$\GZ$ [\GeV{}] & $2.4952 \pm0.0023\pz$
                & $2.4939 \pm0.0024\pz$ & 2.4962$\pz$   & $-0.4$ \\
$\sighadpole$ [nb]   & $41.540 \pm0.037\pzz$ & 
                  $41.491 \pm0.058\pzz$ & 41.480$\pzz$  & $1.6$ \\
$\Rl$          & $20.767 \pm0.025\pzz$ & 
                  $20.765 \pm0.026\pzz$ & 20.740$\pzz$ & $1.1$ \\
$\AFBpole$     & $0.0171 \pm0.0010\pz$ & 
                  $0.0169 \pm0.0010\pz$ & $ 0.0164\pz$ & $0.8$ \\
                                                &&&& \\[-3mm]
$\tau$ polarization:                            &&&& \\
$A_{\tau}$         & $0.1439\pm 0.0042\pz$ & 
                  $0.1431\pm 0.0045\pz$ & 0.1480$\pz$ & $-1.0$ \\
$A_{\Pe}$         & $0.1498\pm 0.0048\pz$ &
                  $0.1479\pm 0.0051\pz$ & 0.1480$\pz$ & $ 0.4$ \\
                       &&&& \\[-3mm]
$\qqbar$ charge asym.:                      &&&& \\
$\swsqeffl$    & $0.2321\pm0.0010\pz$ & 
                  $0.2321\pm0.0010\pz$ & 0.23140     & $ 0.7$ \\
                                             &&&& \\[-3mm]
$m_{\PW}$ [\GeV{}]
& $80.427 \pm 0.046 \pzz$&

  $80.37 \pm  0.09 \pzz\pz$ &80.402$\pzz$ & $ 0.5$ \\

&&&& \\[-3mm]

\hline
\underline{SLD} &&&& \\
$\swsqeffl$ ($A_{\ell}$)
& $0.23098\pm0.00026 $   &
  $0.23109\pm0.00029 $   & 0.23140     & $-1.6$ \\
&&&& \\[-3mm]
\hline
\underline{Heavy Flavor } &&&& \\
$\Rb{}$        & $0.21653\pm0.00069$  &  
                   $0.21656\pm0.00074$  & 0.21578     & $ 1.1$ \\
$\Rc{}$        & $0.1709\pm0.0034\pz$ & 
                   $0.1735\pm0.0044\pz$ & 0.1723$\pz$ & $-0.4$ \\
$\AFBb{}$      & $0.0990\pm0.0020\pz$ &
                   $0.0990\pm0.0021\pz$ & 0.1038$\pz$ & $-2.4$ \\
$\AFBc{}$      & $0.0689\pm0.0035\pz$ & 
                   $0.0709\pm0.0044\pz$ & 0.0742$\pz$ & $-1.5$ \\
$A_{\mathrm{b}}$          & $0.922\pm 0.023\pzz$ &
                   $0.867\pm 0.035\pzz$ & 0.935$\pzz$ & $-0.6$ \\
$A_{\mathrm{c}}$          & $0.631\pm 0.026\pzz$ & 
                   $0.647\pm 0.040\pzz$    & 0.668$\pzz$ & $-1.4$ \\
                                             &&&& \\[-3mm]
\hline
 \underline{$\Pp\bar{\Pp}$ and $\nu$N} &&&& \\
$m_{\PW}$ [\GeV{}] 
 & $80.452 \pm 0.062\pzz$& 
  $80.41  \pm 0.09 \pzz\pz$ & 80.402$\pzz$ & $ 0.8$ \\
$\swsq$ 
& $0.2255\pm0.0021\pz$ & 
  $0.2254\pm0.0021\pz$
 & 0.2226$\pz$ &$ 1.2$ \\
$m_{\mathrm{t}}$ [\GeV{}]
& $174.3\pm 5.1\pzz\pzz$ & 
  $173.8\pm 5.0\pzz\pzz$
                                                      & 174.3$\pzz\pzz$
                                                                    & $ 0.0$ \\
\hline
\end{tabular}\end{center}
}
\end{table}

\section{Changes since RADCOR 1998}

\begin{figure}[t!]
\begin{center}
\epsfig{file=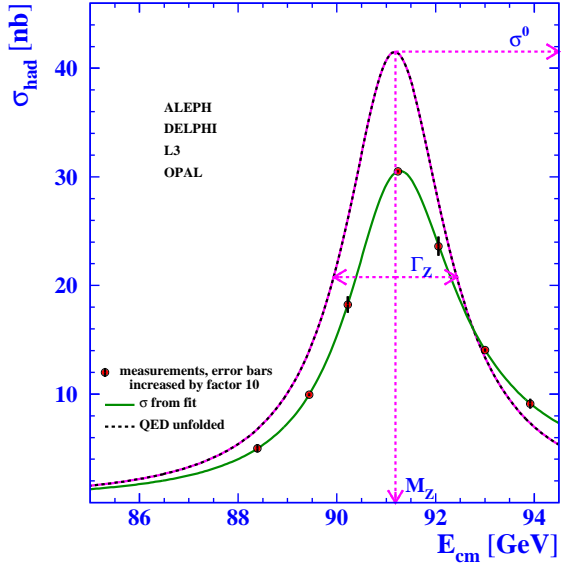,height=3in}
\caption[0]{Illustration of LEP line-shape parameter 
definitions\cite{bib:lepls}.
}
\label{fig:ls}
\end{center}
\end{figure}

The LEP and SLD collaborations present their measurements
in terms of pseudo-observables which are closely related
to the actual measurements,
but include corrections for
effects such as electromagnetic radiation and interference
between photon mediated and $\PZ$ mediated processes.  
These variables, together with other variables commonly
used in electroweak fits are summarized in 
Table~\ref{tab:all}.
The values of the $\PZ$~mass, $\mPZ$, the 
$\PZ$~width, $\GZ$ and the peak hadronic cross section
$\sighadpole$ require quite
large corrections as illustrated in Figure~\ref{fig:ls}
taken from Reference~\cite{bib:lepls}.
Large corrections are also needed to the forward-backward
asymmetries of leptons and quarks.   
The largest corrections of
all are for electron final states, where t-channel
effects dominate in many regions of phase space.
Radiative corrections for the left-right
asymmetries measured by SLD and for polarized 
forward-backward asymmetries are much smaller, but nevertheless
important.  For example, the largest change between
the preliminary SLD measurement of the
left-right asymmetry ($A_{\mathrm{LR}}$)
and the final published value of $A_{\mathrm{LR}}$
came from a correction
to the beam energy which was based on a scan of the $\PZ$
resonance~\cite{bib:sldalr}.

For the five $\PZ$ line-shape parameters (assuming lepton universality)
as measured at LEP\cite{bib:alephls,bib:delphils,bib:l3ls,bib:opalls}, 
the total error and the theoretical
error on the determination of the pseudo-observables is
shown in Table~\ref{tab:th_err}. 
In terms of the effective vector and 
axial and vector couplings of a given fermion to the $\PZ$,
$\gAf$ and $\gVf$:

\begin{eqnarray}
\label{eq:gamma}
\Gammaof{\ff} & = &  \frac{\GF N_{\rm c} \mPZ^3}{6 \pi \sqrt{2} }
        \left(   R_{\mathrm{V}}^{\mathrm{f}} {\gVf}^2 + 
                 R_{\mathrm{A}}^{\mathrm{f}} {\gAf}^2 \right ) + 
\Delta_{\mathrm{QCD}} \\
\label{eq:Af}
A_{\Pf} & = &  2\,\frac{\gVf\gAf}{\gVf^2+\gAf^2} .
\end{eqnarray}
Here, $ R_{\mathrm{V}}^{\mathrm{f}}$ and $R_{\mathrm{A}}^{\mathrm{f}}$
give corrections for final-state QED and QCD effects as well
as quark masses, $\Delta_{\mathrm{QCD}}$ for non-factorizable
QCD effects.  Note that $A_{\Pf}$ depends only on the ratio of 
couplings. 

\def\tableline{\noalign{
\hrule height.7pt depth0pt\vskip3pt}}

\begin{table}[t!]
\caption{\label{tab:th_err}
The total and theoretical errors for the five parameters
in a fit assuming lepton universality.  If lepton universality
is not assumed, the 
theory errors for electrons are larger, i.e.
0.024 for $R_{\Pe}$ and 0.0014$\AFBpole$.
}

\begin{center}
\setlength{\tabcolsep}{9pt}
\renewcommand{\arraystretch}{1.2}
\begin{tabular}{|lcl|ll|}
\hline
Quantity& &  &
Total Error  &  Theory Error \\
\hline

$\mPZ$ & &  
& 2.1 MeV & 0.3 MeV\\
&&&($0.2 \times 10^{-4}$) & $(0.03 \times 10^{-4}$) \\
\hline
$\GZ$  & & 
& 2.3 MeV & 0.2 MeV\\
&&&($9.2 \times 10^{-4}$)  & ($0.8 \times 10^{-4}$)\\
\hline
&&&&\\
$\sigma^0_{\mathrm{had}}$  & $\equiv$ &
$\frac{12 \pi}{\mPZ^2} \frac{ \Gammaof{\epem}\Gammaof{\mathrm{had}}  }
{\GZ^2}$
&  0.037nb & 0.022nb \\
&&&($8.9 \times 10^{-4}$)& ($5.3 \times 10^{-4}$) \\
\hline
&&&&\\
$R_\ell$  & $\equiv$  & 
$\frac{\Gammaof{\mathrm{had}}}{\Gammaof{\ell}}$  
& 0.025 & 0.004\\
&&&($12 \times 10^{-4}$)& ($1.9 \times 10^{-4}$) \\
\hline
$\AFBpole$ & $\equiv$ & 
$\frac{3}{4} A_{\Pe} A_{\Pf}$
& 0.0010 & 0.0001 \\
&&&(5.6\%)&(0.6\%)\\
\hline
\end{tabular}
\end{center}
\end{table}

In almost all
cases the theory
error on the extraction of the pseudo-observables is
at least five times smaller than the experimental error.  
The exceptions
occur for theory corrections involving electron
final states where t-channel photon mediated 
processes are important.  For example, the theoretical error on
luminosity determined with small angle Bhabha scattering 
drives the error on
$\sighadpole$~\cite{bib:lumi,bib:eeee} and 
the effects of interference corrections
on $R_{e} \equiv \frac{\Gammaof{\mathrm{had}}}{\Gammaof{\Pe}}$,
and on ${\AFBe} \equiv \frac{3}{4} A_{\Pe}^2$ give
a theoretical error which is of the same
order as the total error on these quantities when lepton universality
is assumed.

It should be stressed that because of the complexity of the
fitting procedure used to extract the pseudo-observables from 
the several hundred cross section and asymmetries of each of the
LEP experiments, it will be extremely difficult to incorporate
any future improvements in the theory needed to determine the
pseudo-observables.  It is encouraging that, in general,
these theory errors are quite small.  The theory used to extract
the pseudo-observables from the raw measurements has been
very stable since RADCOR 1998 (also shown in Table~\ref{tab:all})
with two exceptions.  
The inclusion of third order
initial state radiation correction shifted 
$\sigma^0_{\mathrm{had}}$  by 0.023 nb or 70\% of its present
total error\cite{bib:thirdqed},  
and also led to 
a $\sim 0.5$~MeV.
shift on $\mPZ$.  Inclusion of initial-state radiation of 
$\epem$ pairs gave rise to a $\sim 0.5 $ MeV shift on $\mPZ$
and $\sim 0.8$ MeV shift on $\GZ$ \cite{bib:pairs}.

\begin{figure}[b!]
\begin{center}
\epsfig{file=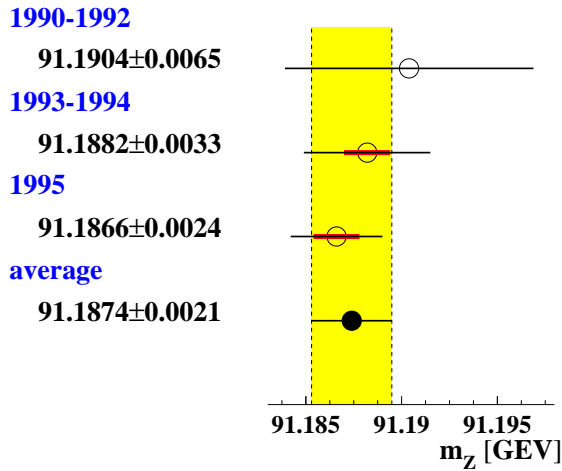,height=3in}
\caption[0]{The value of $\mPZ$ determined in separate 
running periods of the LEP accelerator.
}
\label{fig:mz_abc}
\end{center}
\end{figure}

The largest change in the experimental handling of the data
was due to improved treatment of the errors on the determination
of the beam energy.  A lower energy systematic error 
was obtained for the 1995 LEP run than for 
the 1993 LEP run\cite{bib:lepe}.  
To properly take this into account,
the four experiments combined should give more weight to the
1995 data than each do individually.  To test the 
effects of this reweighting,
separate values of $\mPZ$ were determined for each year as
shown in Figure~\ref{fig:mz_abc}.  Because of the consistency
of $\mPZ$ for the different periods, the numerical effect
of this new procedure was small\cite{bib:lepls}.

In terms of the parameters derived from the LEP line shape, 
the largest change between
RADCOR 1998 and these results is 
that the ratio of the invisible width to the width for
a single generation of charged leptons, 
$\frac{\Gammaof{\mathrm{inv}}}{\Gammaof{\ell}}$,
is now slightly less than the Standard Model prediction
giving a value for the number of neutrinos,
$$
N_{\nu} = 2.984 \pm  0.008,
$$
approximately 2 $\sigma$ smaller than expected.
In 1998, the $N_{\nu}$ value was almost exactly 3.
The change in the central value is 
largely due to an improved
treatment of initial-state radiation that changed the value 
of $\sighadpole$.  The error has also been significantly
reduced because of a reduction in the luminosity theoretical
error~\cite{bib:lumi} since RADCOR 1998.

As is apparent from Table~\ref{tab:all} there have
been big improvements in the heavy quark measurements
made by SLD.  These are discussed in detail in 
Section~\ref{sec:hq}.

\section{Theoretical errors in the determination of $\alpha_s$}

The LEP line-shape data allows a precise determination
of $\alphas$ from the effect of QCD final state corrections
to the hadronic width.  To lowest order, the hadronic
width scales as  
$ \Gammaof{\mathrm{had}} \propto 1 + \frac{\alpha_s}{\pi}$  
which leads
to the following dependences of LEP pseudo-observables
on $\alphas$:
$$\begin{array}{lclcl}
\GZ                     & = & 
\Gammaof{\mathrm{had}} + 3 \Gammaof{\ell} + 3 \Gammaof{\nu} & \propto & 1 + 0.7 \frac{\alphas}{\pi} \\
R_{\ell}  & = & 
\frac{ \Gammaof{\mathrm{had}}}{\Gammaof{\ell}}
& \propto & 1 + \frac{\alphas}{\pi}     \\
\sigma^0_{\mathrm{had}} & = & 
\frac{12 \pi}{\mPZ^2} \frac{ \Gammaof{\ell} \Gammaof{\mathrm{had}}  }
                           {\GZ^2}
& \propto  &  1 - 0.4 \frac{\alphas}{\pi} \\
\sigma^0_{\ell} & = & 
\frac{12 \pi}{\mPZ^2} \frac{ \Gammaof{\ell}^2  }
                           { \GZ^2 }
& \propto  &  1 - 1.4 \frac{\alphas}{\pi} \\
\end{array}$$
Here charged lepton universality has been assumed and the 
invisible width of the $\PZ$ has been assumed to consist of
the contribution from the three generations of 
neutrinos ($3 \Gammaof{\nu}$).  Theoretically, the dependence
of these quantities on $\alphas$ comes from the same
contribution to $\Gammaof{\mathrm{had}}$.  (There are additional
small top mass  ($\mt$) 
contributions to the partial widths from $\alphas \mt^2$ and
$\alphas^2\mt^2$ corrections to the
couplings $\gAf$ and $\gVf$.  The uncertainty on these
corrections correspond to less than a 1.0 GeV uncertainty in the 
top mass and can therefore be ignored.  This is discussed in more
detail below.)
The cleanest measurements of $\alphas$ come from quantities
which depend on the ratio of partial widths such as 
$\Rl$, $\sigma^0_{\mathrm{had}}$  and $\sigma^0_{\ell}$
where additional uncertainties from $\mt$ and 
from the Higgs mass ($\mh$) cancel.

The parameters $\Rl$, $\sigma^0_{\mathrm{had}}$ , and $\sigma^0_{\ell}$
are not independent since
$\Rl = \frac{\sigma^0_{\mathrm{had}}}{\sigma^0_{\ell}}$. 
The LEP parameter set includes  $\Rl$ and 
$\sigma^0_{\mathrm{had}}$, hiding the constraint
imposed by $\sigma^0_{\ell}$ in the correlation
matrix.  The value of $\alphas$ determined
using $\Rl$ alone is (for $\mh = 100$\,GeV) 
$$\alpha_s(\mPZ) =    0.1228 \pm 0.0038 
\ ^{+0.0033 (\mh = 900\,\GeV)}_{-0.0000\,(\mh = 100\,\GeV)}$$
which can be compared to that obtained
from ${\sigma^0_{\ell}}$ alone
$$ \alpha_s(\mPZ)  =      0.1183 \pm 0.0030  
\ ^{+0.0026 (\mh = 900\,\GeV)}_{-0.0000 (\mh = 100\,\GeV)} $$ 
The discrepancy between these two values is another
manifestation of the small value of $N_\nu$ determined 
from these data.
The result of the grand electroweak fit that uses
other electroweak data to constrain the
unknown Higgs mass and includes information from
$\GZ$ is
$$\alpha_s(\mPZ) =  0.1183 \pm 0.0027.$$
The error does not include any theoretical
error from the QCD calculation of  $\Gammaof{\mathrm{had}}$.

The line-shape value of $\alpha_s(\mPZ)$ is comparable to
the recent world averages, such as the PDG average\cite{bib:PDG2000},
$\alpha_s(\mPZ) = 0.1181 \pm 0.002$  and an average of
measurements based on
NNLO calculations\cite{bib:siggi} $\alpha_s(\mPZ) =  0.1178 \pm 0.0034 $.
There is some controversy concerning the theoretical error
for the line-shape $\alpha_s(\mPZ)$.  Values in the literature
differ by nearly an order of magnitude
ranging from 0.0005~\cite{bib:soper} to 0.003~\cite{bib:siggi}.
Given the statistical precision of the line-shape  $\alpha_s(\mPZ)$
measurement it is worth examining the errors in some detail.

The most complete analysis of the error on $\alpha_s(\mPZ)$
is given in Reference~\cite{bib:kuhn} which gives a detailed
examination of the QCD calculations presently implemented
in the commonly used programs TOPAZ0\cite{bib:TOPAZ0} and 
ZFITTER\cite{bib:ZFITTER}.  
The treatment here closely follows that of Reference~\cite{bib:kuhn},
however, calculations which were not available at the time
that this work was completed are also considered here.
The effects of QCD on $\Ghad$
can be divided
into 4 different  categories:  the dominant non-singlet terms which
have the same effect on axial and vector neutral currents;
corrections due to quark masses, dominated by uncertainties
in the b-quark mass;  singlet contributions and finally
propagator corrections associated with the top mass $\mt$.

\subsection{Uncertainties in massless non-singlet terms}
\label{sec:massless}
The non-singlet axial and vector QCD corrections for 
massless quarks in $R_{\mathrm{V}}^{\mathrm{f}}$
and $R_{\mathrm{A}}^{\mathrm{f}}$ are equal
(see Equation~\ref{eq:gamma}).
This correction is known to third order and is given\cite{bib:soper} by:
\begin{equation}
\label{eq-massless}
1 +           \frac{\alphas}{\pi} 
  + 1.40932 ( \frac{\alphas}{\pi} )^2
 - 12.76706 ( \frac{\alphas}{\pi} )^3.           
\end{equation}

One way to assess the errors due to missing
higher order terms is by changing the
QCD renormalization scale, which is explained in some
detail in Reference~\cite{bib:siggi}.  
In  Reference~\cite{bib:soper} the renormalization
scale $\mu$ is varied in the interval 
$ e^{-2} (0.14) < \mu/m_{\mathrm{Z}} < e^2 (7.4)$.
The total fractional
variation of $\Ghad$ for this range of renormalization
scales corresponds
to a variation in
$\alphas$ of 2.6\%, giving errors of approximately $\pm1.3\%$.
A similar study has been done in Reference~\cite{bib:kuhn} and
a total variation in $\alphas$ of 1.0\% is obtained
when $\mu$ is varied in the smaller interval $0.5 < \mu/m_{\mathrm{Z}} < 2.0$.

An alternative method, employed in Reference~\cite{bib:kuhn},
is simply to estimate the possible error due to missing higher 
orders as equal to the last evaluated term.  When this is applied to
Equation~\ref{eq-massless}, the cubic term corresponds to an estimated
error on $\alphas$ of $\pm 1.8\%$.

These uncertainties can be reduced by summing a large class
of ``$\pi^2$-terms'', as is done in Reference~\cite{bib:soper}.
For the measurement of  $\alphas$ from the Z line shape,
considering scale variations
($ e^{-2} (0.14) < \mu/m_{\mathrm{Z}} < e^2 (7.4)$)
and scheme dependence, Reference~\cite{bib:soper} suggests an
error $\pm0.4$\% from QCD theory and that the value of $\alphas$
be scaled by 
\begin{equation}
\label{eq:improve}
{\alphas}^{\mathrm{improved}} = 1.006 {\alphas}^{\mathrm{ZFITTER}}.
\end{equation}
In Reference~\cite{bib:raczka} a similar technique is applied
to $\epem \rightarrow \mathrm{hadrons}$ at $\roots = 31.6$GeV.
In this case the improved value of $\alphas$ was 0.8\%
greater than the standard one, in agreement with Equation~\ref{eq:improve}.
Reference~\cite{bib:raczka} does criticize Reference~\cite{bib:soper} for
not having varied the scheme dependence sufficiently, but this would
appear to be a technical objection as opposed to a practical one,
as the variation in the scheme dependence made only a small 
contribution to the error.

The main controversy surrounding the $\alphas$ error centers
on the validity of the summation of the higher order
terms such as done in Reference~\cite{bib:soper}.
This summation has also been applied to the
determination of $\alphas$ using information
from hadronic decays of the
tau lepton, $R_{\tau}$.
Given the much larger value of $\alphas$ at 
this scale, it might be expected that any problems in
the summation procedure would be amplified.  Using 
a method called Contour Improved 
Fixed Order Perturbation Theory (CIPT)
\cite{bib:pich},
which is
similar to the $\pi^2$ summation of Reference~\cite{bib:soper},
OPAL\cite{bib:opaltau} obtains 
$\alphas (m^2_{\tau}) = 0.348 \pm 0.010 (\mathrm{exp}) \pm 0.019 
(\mathrm{theory})$.
However, it has also been found that when ``renormalon'' effects
refered to as ``renormalon chain resummation'' 
(RCPT)~\cite{bib:braum}
are included, 
$\alphas (m^2_{\tau}) 0.306 \pm 0.005 (\mathrm{exp}) \pm 0.011
(\mathrm{theory})$
is obtained.  

Since these two methods of determining $\alphas$ with
$R_{\tau}$ marginally disagree by more than the theoretical
error estimates, it is important that an error due to
renormalon effects be included in any analysis
of $\alphas$ from the $\PZ$ line shape. 
Fortunately, the effects of renormalons were also 
included in the calculations 
of Reference~\cite{bib:soper} for
$\Ghad$ and were found to have almost no effect.  
	
Since these renormalon effects 
are small at the $\PZ$ resonance, 
I conclude that 
the studies of $R_{\tau}$ give no evidence for additional
QCD uncertainties in the non-singlet term and the 
0.4\% error estimate is appropriate.
Of course, it cannot be excluded that there are other unknown effects,
but this is true for all of the theoretical errors
in the $\PZ$ resonance studies as well.

\subsection{Mass correction}
\label{sec:mass}

The only significant contribution from the uncertainty in the
mass corrections is from the b-quark mass.  The uncertainties
associated with these corrections can be significantly reduced
if the running b-quark mass, $\overline{m}_b(\mPZ)  \simeq 2.77$ GeV
is used.

These corrections are known to ${\cal O}(\alphas^3)$ for
$R_{\mathrm{V}}$ (vector current),
but only to ${\cal O}(\alphas^2)$ for 
$R_{\mathrm{A}}$ (axial-vector current). In Reference~\cite{bib:kuhn}
uncertainties from missing orders are evaluated using
scale variations  ($0.5 < \mu/\mPZ < 2.0$) 
and from the size of the  ${\cal O}(\alphas^2)$ terms.  
The scale variation gave a total variation in $\alphas$ of
0.04\%.  The size of the axial-vector ${\cal O}(\alphas^2)$ term
dominates the uncertainty, and corresponds to 0.05\%.
I adopt $\pm 0.05\%$ as the error estimate from unknown
higher orders in the mass  corrections.

Propagating the error of the pole mass of the b-quark,
$M_b = 4.7 \pm 0.2$~GeV, an error on $\alphas(\mPZ)$
of $\pm0.31\%$ is obtained.

\subsection{Singlet contributions}
\label{sec:singlet}

The error on the singlet contribution is due to
uncertainties from the top quark mass and from possible
missing higher orders.  The error due to the top-quark
mass is evaluated using ZFITTER or TOPAZ0 and is not included
in the QCD error estimate.  
The QCD singlet contribution
(including top mass dependent contributions with
$m_t = 174$~GeV) scales $\Ghad$ by
\begin{equation}
\label{eq:singlet}
1 - 0.63 (\alphas/\pi)^2 - 2.69 (\alphas/\pi)^3 .
\end{equation}
Varying the renormalization scale in the
range ($0.5 < \mu/\mPZ < 2.0$) in this expression gives a total variation
of $\alphas$ of 0.26\%, whereas dropping the third term
changes $\alphas$ by 0.38\%.
The larger value is taken as the error.

\subsection{Propagator corrections}
\label{sec:propagator}

The behavior of the widths themselves, such as $\GZ$ or $\Ghad$,
will differ from observables
that depend on the ratio of widths, $\Rl$, $\sigma_{\mathrm{had}}^0$ and
$\sigma_{\ell}^0$.  Since the $\alphas$ correction to the propagator 
affects all partial widths equally, its effects will cancel
in these observables.
This correction is parameterized\cite{bib:kuhn-rho} by
\begin{equation}
\delta \rho_{\mt} = \frac{3 \sqrt{2} \GF \mt^2}{16 \pi^2} 
( 1 - 2.8599 \frac{\alphas}{\pi} - 14.594 (\frac{\alphas}{\pi})^2 )
\end{equation}
where $\mt$ is the top pole mass and $\alphas = \alphas(\mt) \simeq 0.11$.
For $\mt = 175$ GeV the partial widths of leptonic and neutrino
final states scale with $\alphas$ as 
\begin{equation}
1 + \delta \rho_{\mt}^{\mathrm{QCD}} = 
1 - 0.027\frac{\alphas}{\pi}   - 0.140(\frac{\alphas}{\pi})^2 .
\end{equation}
Although this propagator correction is only a few percent of the
QCD hadronic final-state correction, its theoretical uncertainty can be
disproportionately large.
Measurements of $\alphas$ through quantities in which the propagator effects
cancel, such as $\Rl$, $\sigma^0_{\mathrm{had}}$ and $\sigma_{\ell}^0$ are therefore favored.
Such observables also benefit from the cancelation 
of other effects in the propagator,
such as $\mh$ and $\mt$ dependence, which are in fact much larger than the
QCD effects.

The scheme and renormalization dependence of the QCD propagator
correction has been evaluated in Reference~\cite{bib:kuhn-rho}
and is less than $5 \times 10^{-5}$.  
%
%
Taking into account the  Z branching ratio
to hadrons ($\sim 70$\%),
the corresponding additional
error on $\alphas$ determined from $\GZ$ is 0.21\%.  

This renormalization scale uncertainty 
could also be viewed as an error on $\mt$. 
For $\mt = 175$ GeV, this corresponds to an uncertainty
of 0.4 GeV, which is much smaller than the corresponding
experimental uncertainty of 5 GeV.

In the extreme alternative approach of
taking the last calculated term as the error estimate, a
fractional error on $\GZ$ of $17 \times 10^{-5}$ is obtained,
corresponding to an additional error on $\alphas$ determined
from $\GZ$ of 0.7\%.

Effects from the uncertainties of the QCD corrections
(also known to second order)
on the ratio of couplings for 
different fermions, $\gVf/\gAf$, or equivalently 
$\swsqweff$, have been justifiably neglected\cite{bib:kuhn-kappa}
in these error estimates.
These corrections give rise to
slight differences in
the $\alphas$ dependence of the Standard Model predictions for
$\Gammaof{\nu}$ and $\Gammaof{\ell}$.

\def\tableline{\noalign{
\hrule height.7pt depth0pt\vskip3pt}}
\begin{table}[t!]
\caption{\label{tab:errsum}
Summary of the QCD error on $\alphas$ derived from various line shape
observables based on the ratio of partial widths, 
$\Rl$, $\sigma_{\mathrm{had}}^0$ and $\sigma_{\ell}^0$
and from the total width of the $\PZ$, $\GZ$.
}
\begin{center}
\begin{tabular}{|l|cc|}
\hline
effect   & $\alphas$ from ratio of widths & $\alphas$ from $\GZ$ \\
\hline
missing orders, massless, non-singlet   &  0.40\% & 0.40\% \\
missing orders, singlet                 &  0.38\% & 0.38\% \\
missing orders, mass                    &  0.05\% & 0.05\% \\
b-quark mass                            &  0.31\% & 0.31\% \\
propagator effects                      &     -   & 0.21\% \\
\hline
Total                                   &  0.84\% & 0.87\% \\
\hline
\end{tabular}
\end{center}
\end{table}

\subsection{$\alphas$ summary} 

The effects discussed above are summarized in
Table~\ref{tab:errsum}.  
I conservatively assume that the singlet and non-singlet
contributions could be 100\% correlated and sum their
errors linearly.  
Another approach to possible correlation
between theoretical uncertainties in different parts of the
QCD calculation is taken by Reference~\cite{bib:siggi}. 
This approach is based on an attempt to extract the overall
dependence of $\Rl$ on $\alphas$, including all contributions
to the running quark masses and any residual propagator effects
from a third order fit in $\alphas$ 
to the ZFITTER output as a function of $\alphas$~\cite{bib:tournefier}.
Note that because of the running quark masses and the propagator 
corrections the expansion contains terms beyond the third-order.
The effects of these terms are
included by the fit in the effective coefficients of the
lower-order terms.
The renormalization group equations are then applied to the resulting
expansion.   An error from 
renormalization scale uncertainties of +2.4\%, -0.3\% is
obtained which is compatible to the result one obtains
from adding the errors of the singlet and non-singlet 
contributions {\em without} the
correction of Equation~\ref{eq:improve}.
Note that neither ZFITTER nor TOPAZ0 presently
include this correction.

The other effects in Table~\ref{tab:errsum} 
are added in quadrature.
The contribution of the QCD uncertainty in the measurement of
$\alphas$ from $\Rl$, $\sigma_{\mathrm{had}}^0$ and $\sigma_{\ell}^0$
is slightly smaller than that from $\GZ$ (or the derived quantity $\Ghad$)
because the propagator corrections are smaller.  
Since the
constraints on $\mt$ and $\mh$ are presently much looser than
the uncertainty on the QCD effect in the propagator, this 
additional source of error could be ignored in the grand electroweak
fit.  However, at present this is numerically unimportant.
Rounding
the QCD theoretical uncertainty to 0.9\% and applying the correction
of Equation~\ref{eq:improve} gives an improved value
of the strong coupling constant
$$\alpha_s(\mPZ) =  0.1190 \pm 0.0027(\mathrm{Exp.+EW}) \pm 
0.0011 (\mathrm{QCD}) $$
where the first error includes statistical, systematic and electroweak errors
and the second error is due to QCD effects.  This result
is nearly as precise as the 2000 PDG\cite{bib:PDG2000} world average 
of $\alpha_s(\mPZ) = 0.1181 \pm 0.002$.

\section{Measurements of $\swsqeffl$ and $A_b$ }
\label{sec:hq}

\begin{figure}[t!]
\begin{center}
\epsfig{file=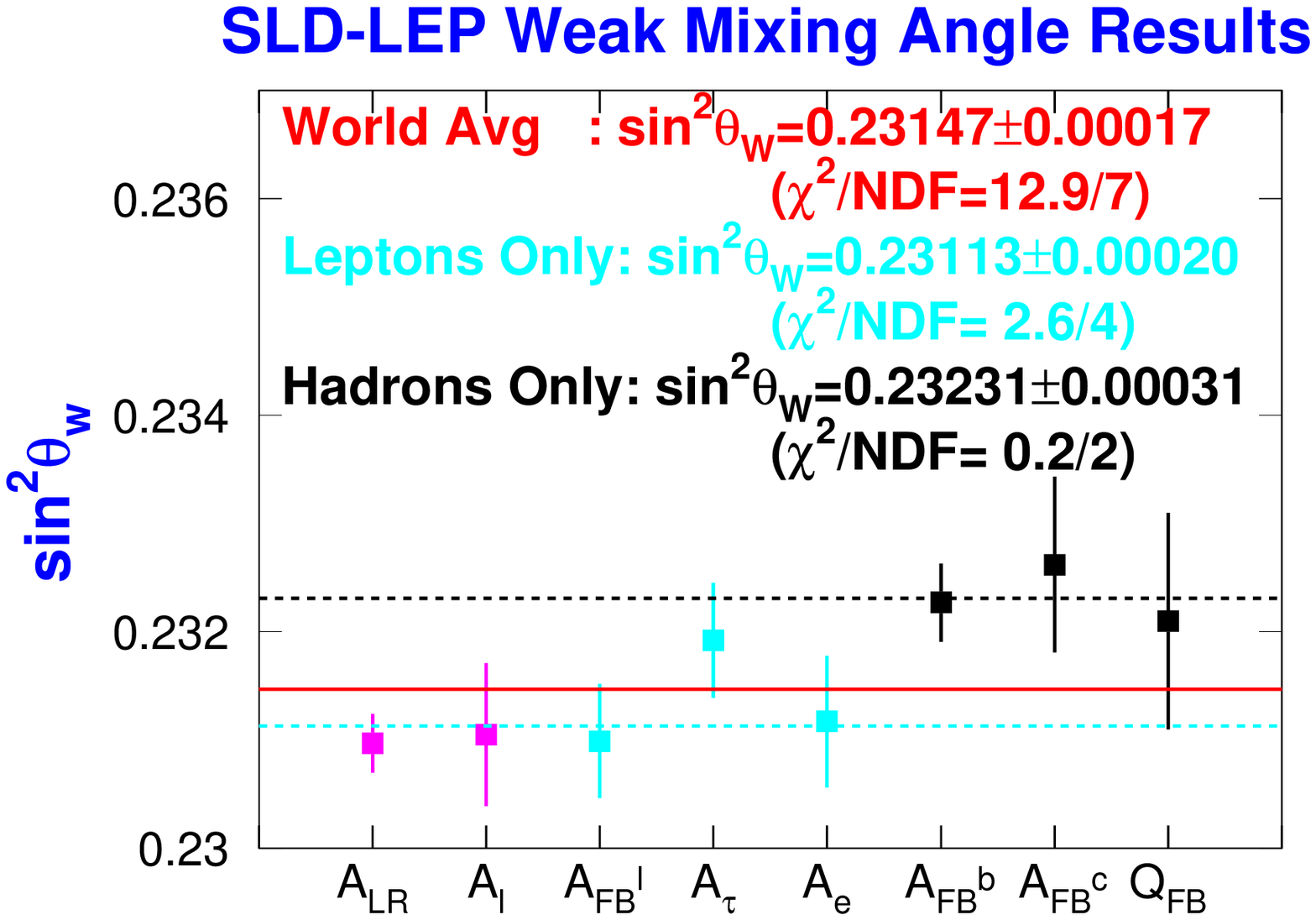,height=4in}
\caption[0]{Comparison of $\swsqeffl$    measured at SLC and LEP\cite{bib:toshi}.
}
\label{fig:sinsq}
\end{center}
\end{figure}

The effective value of the weak mixing angle is given
by ratio of effective vector and axial couplings
\begin{equation}
\swsqeffl \equiv \frac{1}{4}  (1 - \frac{\gVl}{\gAl})
\end{equation}
and is closely related to $A_\ell$ as given by Equation~\ref{eq:Af}.
The most accurate value of $\swsqeffl$ comes
from the
left-right asymmetry
$A_{\mathrm{LR}} = A_{\mathrm{\ell}}$ as measured by SLD\cite{bib:sldalr}.  
Additional constraints
come from polarized forward-backward asymmetries of leptons
measured at SLD and from forward-backward asymmetries measured
at LEP (see Table~\ref{tab:th_err}).  LEP can also probe $A_{\Pe}$
and $A_{\tau}$ directly using the tau polarization.  The values
presented here include an improved preliminary
measurement from OPAL~\cite{bib:opaltaup} and a recent
final result from DELPHI~\cite{bib:delphitaup}.
The resulting
values of $\swsqeffl$  are given in Figure~\ref{fig:sinsq}.

The forward-backward asymmetries of $\bbbar$ and $\ccbar$ events can
also be used to determine  $\swsqeffl$  through the relations
$\AFBc  \equiv \frac{3}{4} A_{\Pe} A_{\mathrm{c}}$ and
$\AFBb  \equiv \frac{3}{4} A_{\Pe} A_{\mathrm{b}}$ 
as long as the
Standard Model is used to calculate  $A_{\mathrm{c}}$  and $A_{\mathrm{b}}$.
The figure shows that there is a significant discrepancy between
the quark based forward-backward measurements and those made in the
leptonic sector alone.

Since the leptonic measurements agree well,
a possible explanation for this effect would be that the values
of either or both $A_{\mathrm{c}}$ and $A_{\mathrm{b}}$ 
deviate from the Standard Model prediction.  Using the polarized
forward-backward asymmetry for $\bbbar$ and $\ccbar$ events
$A_{\mathrm{c}}$ and $A_{\mathrm{b}}$ can be obtained
directly~\cite{bib:sldhq}.
The 
comparison of the SLD result and Standard Model is given 
in Table~\ref{tab:abc}. 
The precision of the preliminary 
SLD result has been considerably improved
since the 1998 RADCOR~\cite{bib:sldhq}, (see Table~\ref{tab:all}
and Reference~\cite{bib:lepew2000}) but
the SLD data are not statistically precise enough to indict the Standard Model
by themselves.  It is unfortunate that SLD was prevented from
running long enough to settle this issue.

Given that the SLD data cannot confirm a deviation in the values of
$A_{\mathrm{c}}$ and $A_{\mathrm{b}}$, the remaining possibilities are
a large statistical fluctuation or an unstated systematic error in
some of the measurements.  The most economical solution would be a
systematic error affecting the LEP heavy quark measurements.  However,
there is no obvious source of such an error. 
On the experimental
side, the total systematic error for $\AFBb$ would have to be inflated
by more than a factor of 5 to account for the discrepancy.  The
theoretical systematics are dominated by the correction to the
observed asymmetry for gluon radiation.  
For a completely inclusive selection, 
the total QCD correction 
applied to the data is approximately 4\%, of order the discrepancy
between the LEP $\AFBb$ average and the expected value from the
Standard Model.  However, most experimental analyses tend to 
reject  events strongly affected by gluon radiation so the
actual corrections are much smaller~\cite{bib:Abbaneo}.
Furthermore, the experimental techniques used in jet-charged
measurements uses data driven correction which attenuate
the QCD effects still further.
The residual QCD error on the LEP $\AFBb$ measurements,
largely from missing higher orders in the QCD calculation
and from hadronization, is estimated to be 0.2\%.

The LEP heavy quark results are not all published or final
and it is expected that
some of the techniques developed for b-mixing studies will
result in an increase in the precision of some of the LEP
results~\cite{bib:delphihq,bib:alephhq}.

\def\tableline{\noalign{
\hrule height.7pt depth0pt\vskip3pt}}
\begin{table}[t!]
\caption{\label{tab:abc}
Comparison of measured and Standard Model values of
$A_{\mathrm{c}}$ and $A_{\mathrm{b}}$.  The LEP
values are extracted using the LEP/SLD average
of  $A_{\mathrm{\ell}} = 0.1500 \pm 0.0016$
}
\begin{center}
\setlength{\tabcolsep}{9pt}
\renewcommand{\arraystretch}{1.2}
\begin{tabular}{|l|ll|}
\hline
                & $A_{\mathrm{b}}$            & $A_{\mathrm{c}}$  \\
\hline
SLD             & $0.922 \pm 0.023$  (-0.6 $\sigma$)        &  
                  $0.631 \pm 0.026$  (-1.4 $\sigma$) \\
LEP/SLD Average & $0.898 \pm 0.015$  (-2.5 $\sigma$)         &  
                  $0.623 \pm 0.020$ ( -2.2 $\sigma$) \\
\hline
Standard Model   &  0.935                      &  
                   0.668        \\
\hline
\end{tabular}

\end{center}
\end{table}

\section{Impact of uncertainties in hadronic vacuum polarization}

The constraint given by the LEP and SLD asymmetry data on
the Higgs mass is
strongly dependent on the
running value of $\alpha$ parameterized by
$$\alpha({\mPZ}^2) = \frac{ \alpha(0)}
{ 1 - \Delta \alpha_{\ell}({\mPZ}^2) 
    - \Delta \alpha_{\mathrm{had}}^5 ({\mPZ}^2) 
    - \Delta \alpha_{\mathrm{top}} ({\mPZ}^2)   }$$
where $\Delta \alpha_{\ell}({\mPZ}^2)$ is the contribution
to vacuum polarization from leptons,  
$\Delta \alpha_{\mathrm{top}} ({\mPZ}^2)$ the contribution
from top quarks and  $\Delta \alpha_{\mathrm{had}}^5 ({\mPZ}^2)$
the contribution from the five lightest quarks.  
The value of  $\Delta \alpha_{\mathrm{had}}^5 ({\mPZ}^2)$
is derived from the measured cross section for
the process $\epem \rightarrow \mathrm{hadrons}$ at low
energies
and currently limits the precision with which $\alpha({\mPZ}^2)$
can be determined. 

\begin{figure}[b!]
\begin{center}
\epsfig{file=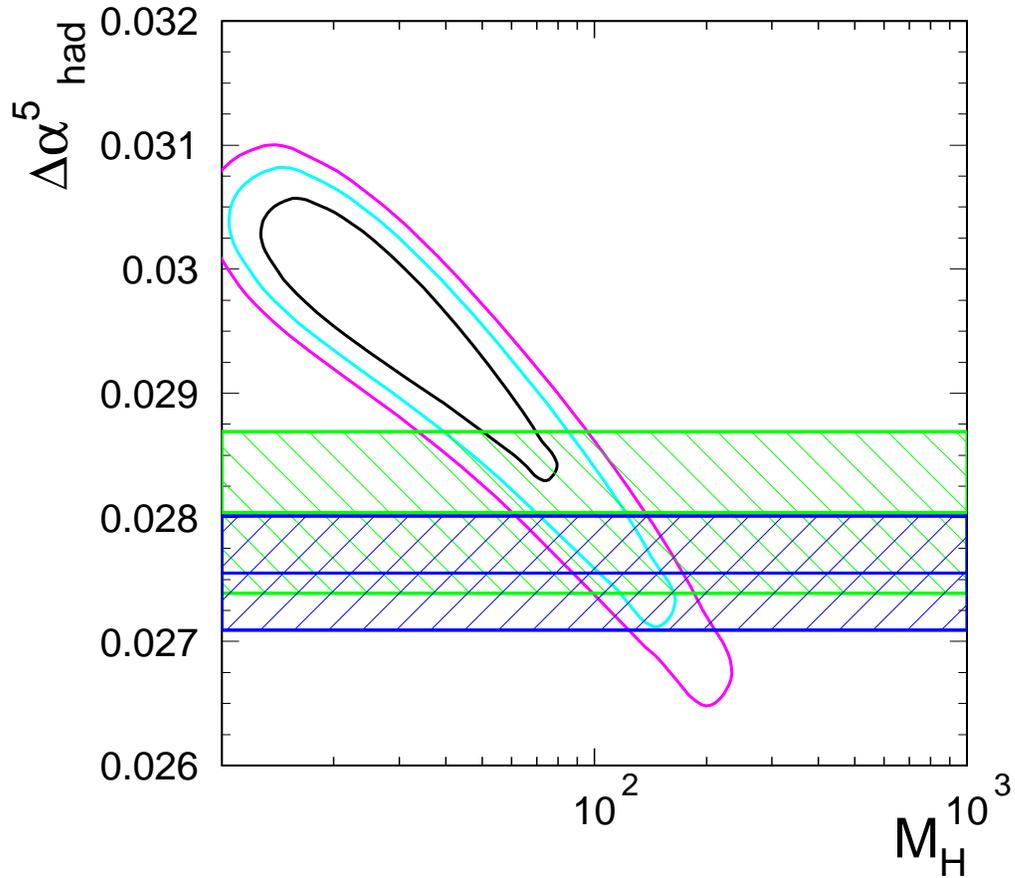,height=6in}
\caption[0]{ The contours show the 1 $\sigma$ (47\% C.L.), 
2 $\sigma$ (91\% C.L.) and 3$\sigma$ (99.5\% C.L.) limits in the 
$\Delta \alpha_{\mathrm{had}}^5 ({\mPZ}^2)$-$\mh$ plane,
for a data similar, 
but not indentical to that of Table~\ref{tab:all}\cite{bib:toshi}. 
The upper bands show the 
value from $\Delta \alpha_{\mathrm{had}}^5 ({\mPZ}^2)$-$\mh$
from Reference~\cite{bib:ej}
and the lower band shows preliminary results using the new
preliminary BES data from Reference~\cite{bib:pietrzyk}
\label{fig:aemh}}
\end{center}
\end{figure}

\begin{figure}[t!]
\begin{center}
\epsfig{file=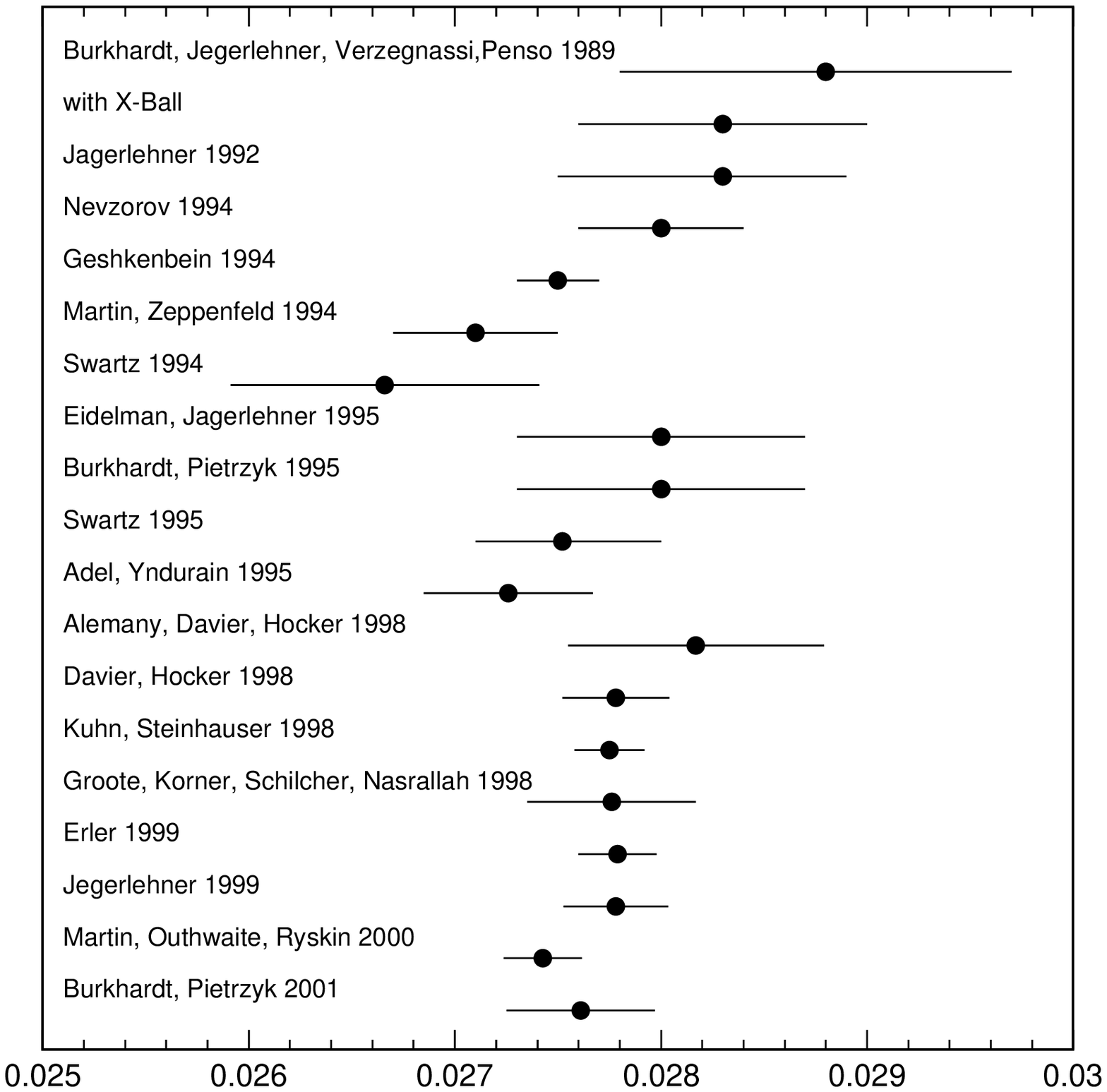,height=6in}
\caption[0]{Compilation of values of 
$\Delta \alpha_{\mathrm{had}}^5 ({\mPZ}^2)$ as a function of
time. (See References~\cite{bib:ej,bib:burkhardt,bib:alemany,bib:davier,bib:kuhn-alpha,bib:groote,bib:erler,bib:jegerlehner,bib:martin},
for older values see citation in Reference~\cite{bib:burkhardt}).
The last two results incorportate the new BES data\cite{bib:BES}.

\label{fig:aemevo}}
\end{center}
\end{figure}

The correlation between  $\Delta \alpha_{\mathrm{had}}^5 ({\mPZ}^2)$
and the determination of the Higgs mass from the electroweak
data is shown in Figure~\ref{fig:aemh}.  The value of the
Higgs mass determined from the fit is strongly correlated
with the  $\Delta \alpha_{\mathrm{had}}^5 ({\mPZ}^2)$ input.
In Figure~\ref{fig:aemevo} various determinations
~\cite{bib:ej,bib:burkhardt,bib:alemany,bib:davier,bib:kuhn-alpha,bib:groote,bib:erler,bib:jegerlehner,bib:martin,bib:pietrzyk} 
 of the
$\Delta \alpha_{\mathrm{had}}^5 ({\mPZ}^2)$ are shown.  The
LEP Electroweak group has generally used the value from
Eidelmann and Jegerlehner~\cite{bib:ej} which is primarily
data driven. It is interesting that the new determination
from Pietrzyk~\cite{bib:pietrzyk}, 
based on new data from BES~\cite{bib:BES} presented at ICHEP 2000 
agrees well with the result of theory driven
results
~\cite{bib:alemany,bib:davier,bib:kuhn-alpha,bib:groote,bib:erler,bib:jegerlehner,bib:martin} 
which makes use of perturbative QCD.  In any case the result of 
using the Pietrzyk result is to move the Higgs mass prediction of the
grand electroweak fit towards higher values (see Table~\ref{tab:higgs}).
We can expect that the error on $\Delta \alpha_{\mathrm{had}}^5 ({\mPZ}^2)$
will continue to decline in the future as more data is collected by BES
and other low energy electron-positron storage rings.

\begin{table}[t!]
\caption[0]{\label{tab:higgs}
Limits and values for Higgs mass determined from fits with 
the ``usual'' experiment driven value\cite{bib:ej} of 
$\Delta \alpha_{\mathrm{had}}^5 ({\mPZ}^2)$ traditionally
used by the LEP electroweak group and the
value of $\Delta \alpha_{\mathrm{had}}^5 ({\mPZ}^2)$
presented at ICHEP2000 including BES data~\cite{bib:pietrzyk}.
}
\begin{center}
\begin{tabular}{|l||l|l|}
\hline
$\Delta \alpha_{\mathrm{had}}^5$ & 
$0.02804 \pm 0.00065$  &
$0.02755 \pm 0.00046$ \\ 

\hline
&&\\
$\mh$ &
$60 ^{+52}_{-29} $~GeV &
$90 ^{+63}_{-39} $~GeV \\
$\mh$ 95\% C.L. upper limit   &
$ \mHiggs < 170 \mbox{GeV} $  &
$ \mHiggs < 210 \mbox{GeV} $   \\
\hline
\end{tabular}
\end{center}
\end{table}

\clearpage
\section{Conclusions}

The LEP results for the $\PZ$ line shape and lepton asymmetries have been
stable for some time and are now final and published\cite{bib:lepls}.  The 
theoretical
error assigned to
$\alphas$ determined from these data remains controversial.
If the correction of Reference~\cite{bib:soper} is applied, 
$$\alpha_s(\mPZ) =  0.1190 \pm 0.0027(\mathrm{Exp.+EW}) \pm 
0.0011 (\mathrm{QCD}) $$
is obtained.  This is competive with  
the 2000 PDG\cite{bib:PDG2000} world average 
of $\alpha_s(\mPZ) = 0.1181 \pm 0.002$.

The SLD measurement of  $\swsqeffl$, based primarily on the left-right
polarized asymmetry, 
is also now final and published\cite{bib:sldalr}.
Its value agrees with that obtained from lepton asymmetries and 
$\tau$ polarization at LEP.  However, the average from these lepton
based results is in disagreement with the LEP heavy-quark 
measurements of $\swsqeffl$.  It is possible that the discrepancy
could be explained by anomalous values of $A_{\mathrm{b}}$ or
$A_{\mathrm{c}}$, but the direct measurements of these quantities
by SLD are in agreement with both the Standard Model and
the LEP values, assuming a Standard Model value for $A_{\Pe}$.

The interpretation of these electroweak results in terms of
limits on the Higgs boson mass depends on the value of
$\Delta \alpha_{\mathrm{Had}}^5 ({\mPZ}^2)$.  New 
$\epem$ cross section measurements from
BES gives a data driven value
$\Delta \alpha_{\mathrm{had}}^5 ({\mPZ}^2)$
which agrees with previous theory driven determinations,
resulting in a higher prediction for the Higgs boson mass.

\Acknowledgments
I would like to thank the ALEPH, DELPHI, L3, OPAL and
SLD collaborations and the LEP and SLD 
Electroweak Working Groups for their assistance in preparing
this report.  I am especially grateful to Martin 
Gr\"{u}newald and G\"{u}nter
Quast for providing the electroweak fits used in this report.
I would also like to thank Dave Soper and Sigi Bethke 
for useful discussions about the application of QCD
to the LEP line-shape results.  
Finally I wish to thank Dick Kellogg for his suggestions and encouragement.

\end{document}